\documentclass[conference]{IEEEtran}
\usepackage[latin9]{inputenc}
\usepackage{url}
\usepackage{amsmath}
\usepackage{amssymb}
\usepackage{graphicx}
\usepackage[bookmarks=false]{hyperref}
\makeatletter

\ifCLASSINFOpdf
\else
\fi
\usepackage[symbol,flushmargin]{footmisc}

\@ifundefined{showcaptionsetup}{}{%
 \PassOptionsToPackage{caption=false}{subfig}}
\usepackage{subfig}
\makeatother

\begin{document}

\title{Deep Learning of Human Perception in Audio Event Classification}

\author{\IEEEauthorblockN{ Yi Yu $^{1,2}$, Samuel Beuret $^{3}$, Donghuo Zeng$^{2,1}$,  Keizo Oyama$^{1,2}$}

\\
\IEEEauthorblockA{ $^1$National Institute of Informatics, Tokyo, $^2$SOKENDAI \\
$^{3}$ \'Ecole Polytechnique F\'ed\'erale de Lausanne\\
}
}

\maketitle

\begin{abstract}

In this paper, we introduce our recent studies on human perception in audio event classification by different deep learning models. In particular, the pre-trained model VGGish is used as feature extractor to process audio data, and DenseNet is trained by and used as feature extractor for our electroencephalography (EEG) data. The correlation between audio stimuli and EEG is learned in a shared space. In the experiments, we
record brain activities (EEG signals) of several subjects while they are listening to music events of 8 audio categories selected from Google AudioSet, using a 16-channel EEG headset with active electrodes. Our experimental results demonstrate that i) audio event classification can be improved by exploiting the power of human perception, and ii) the correlation between audio stimuli and EEG can be learned to complement audio event understanding.
\end{abstract}


\footnotetext{$^{3}$ Samuel was involved  in  this  work  during  their  internship  in National Institute of Informatics (NII), Tokyo.}

\section{Background and Motivation}


Audio event classification is an interesting problem in machine perception, which mainly targets for recognizing and relating sounds from audio. Various Convolutional Neural Networks (CNNs) have demonstrated promising results in audio classification \cite{StoberNIPS14}. In contrast, human perception and responses to audio event, e.g., how to understand and interpret audio events encountered in real-world environments by specific semantic categorization or more detailed description, is correlated to human cognitive processes. Recent researches on cognitive neuroscience \cite{StoberNIPS14}\cite{Francisco17} have been carried out to learn discriminative features from EEG recordings for distinguishing music audio stimuli by CNN techniques, which have shown the possibility of using brain signals and deep learning in classifying music audios. However, little research studies the following problems: i) how to measure the differences between audio events by exploiting audio and/or EEG, and ii) how to measure the correlation between audio stimuli and the corresponding EEG data.

Motivated by Google research which recently released a sound vocabulary and dataset aiming to provide a common, realistic-scale evaluation platform for audio event classification such as human sounds, music genres, environmental sounds (see https://research.google.com/audioset/), in this work, we build a new EEG dataset to annotate Google audio sub-dataset with 160 segments for singing entity. EEG data are recorded while subjects are listening to the selected music audios, for the purpose of annotating audio events. Particularly, alignments between audio and EEG are obtained during EEG data collection. We study not only the capabilities of deep learning in classifying audio stimuli by using the evoked EEG data, but also the correlation between audio feature and EEG feature to understand audio events. This paper has two major contributions: i) several models are compared to evaluate the performances of audio event classification. ii) the correlation between audio feature and EEG feature is learned to help audio event understanding and classification, which generates competitive results.

\begin{figure*}
\subfloat[EEG only]
{\protect\includegraphics[height=4.8cm]{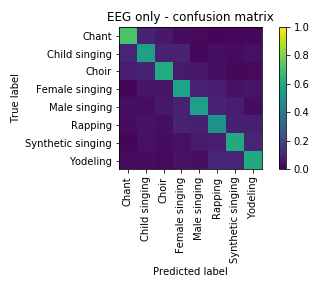}}
\hfill{}
\subfloat[Audio only]
{\protect\includegraphics[height=4.8cm]{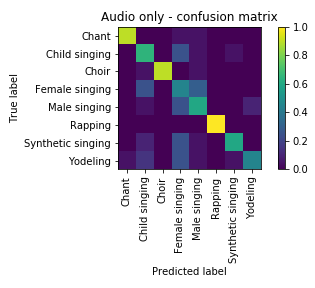}}
\hfill{}
\subfloat[Audio and EEG]
{\protect\includegraphics[height=4.8cm]{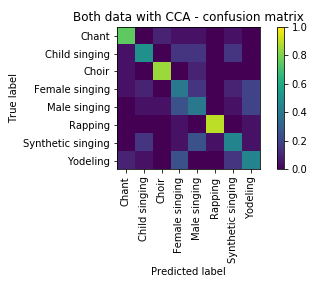}}
\caption{\label{fig:confmax}Confusion matrix for audio event classification.}
\end{figure*}

\section{Methodology}
This paper aims to demonstrate how audio stimuli, EEG, and their combination distinguish different audio events. To this end, we train different deep models to learn the correlation
between audio stimuli and EEG and investigate several scenarios of audio event classification.



\subsection{Audio event classification using EEG data alone}

We first train a convolutional neural network DenseNet \cite{iandola2014densenet}, which takes EEG data as input, has a dense layer as output layer, and uses a softmax activation function for the classification of audio events. We try to minimize the cross-entropy between the predicted probabilities and the reference probabilities (the class labels). The network trained here (without the last dense layer) is reused in all the other learning models as an EEG feature extractor. It allows us to reduce each EEG recording to a feature vector of 512 dimensions.

For the comparison purpose we perform event classification following an equivalent pipeline. The first step is to reduce 512-dimension EEG feature to a dimension of 20 using PCA \cite{jol11}. In the second step, we train a SVM classifier \cite{Hea98} based on the 20-demension compact feature and the class label. 

\subsection{Audio event classification using audio data alone}

The class labels are predicted by solely using the audio data. The first step is to extract the necessary features from the raw audio. This is done by using the pre-trained VGGish model \cite{Her17} which extends the well-established VGG16 architecture \cite{Lon15} trained on the large-scale AudioSet \cite{Gem17}. In this way, each song is reduced to a 1152-dimension vector. This vector is further reduced to a 20-dimension vector using PCA. On this basis, we train a SVM classifier for classifying audio events.

\subsection{Audio event classification using both audio and EEG}

A pair of EEG data and audio signal are reduced to 512 and 1152 dimensions using the previously defined feature extractors. Then, they are concatenated together as a 1664-dimension vector, which is further reduced to 20 dimensions by using PCA. On this basis, a SVM classifier is trained.

\subsection{Correlation learning between audio and EEG}

We use canonical-correlation analysis (CCA) \cite{Hotelling36}, Deep CCA (DCCA) \cite{And13}, and Category-based Deep CCA(C-DCCA) \cite{yiyu18} to project audio and EEG features into a shared space. We hope that the information contained in the EEG data can help extract meaningful features from the audio data through canonical-correlation analysis. To learn the correlation between audio and EEG, two tasks are investigated: using EEG as query to retrieve audio and vice versa.

\section{Evaluation}

Our audio events, selected from Google large-scale AudioSet \cite{Gem17}, contain 8 audio categories (Chant, Child singing, Choir, Female singing, Male singing, Rapping, Synthetic singing, and Yodeling) with 160 10s-long audio segments. We conduct the EEG data collection on 9 male subjects, using EEG devices produced by OpenBCI (\url{http://openbci.com/}) where 16 channels are used to sample EEG data at the frequency of 125 Hz. Each subject listens to a category-based session with 20 audios (pause time between audios is 2 seconds) for 5 times while his EEG signal is recorded. A total number of 7200 EEG signals are acquired. We randomly split our dataset into ten folds, and make sure that each category is equally represented in each fold. Then, all our test results are averaged over 10 cross validations.

\subsection{Comparisons among proposed learning models}

Using DenseNet with a fully-connected layer as a classifier, we obtain a training accuracy of 94\% and a testing accuracy of 61\%. When we use the DenseNet as a feature extractor and a combination of PCA and SVM to perform the classification, we obtain a training accuracy of 98\% and a testing accuracy of 59\%. We can see that the difference between two testing accuracies is small, showing that the classification using PCA and SVM is only slightly lower than that of the dense layer used to train the network. It shows that we can consider the results obtained by this classifier with dense layer on the next experiment as reasonable, compared to an optimal classifier. The confusion matrix associated with this experiment is shown in Fig. \ref{fig:confmax}(a). It shows that the error is more or less equally distributed among all the classes, without any obvious error pattern.

For the classification considering only audio data, we obtained a training accuracy of 100\% and a testing accuracy of 67\%. The confusion matrix associated with the results is shown in Figure \ref{fig:confmax}(b). We can notice some interesting error patterns, such as the frequent confusions between female singing and child singing or female singing and male singing. The observations are consistent with the reality, where the average frequency range of women is situated between those of children and men. For the classification considering both audio and EEG data, we obtain a training accuracy of 99\%, and a testing accuracy of 81\%. This testing accuracy is much higher than that achieved by EEG-only and audio-only methods. It shows that a part of the information contained in two mediums is mutually complementary. The confusion matrix corresponding to this experiment is shown in Figure \ref{fig:confmax}(c). It seems to combine the characteristics of the two previous experiments, although recurrent errors have been attenuated. All results are summarized in Table \ref{tab:res}.

\begin {table}[h]
  \caption {Accuracies of audio event classification with different learning methods and data modalities} \label{tab:res}
  \begin{center}
    \begin{tabular}{| c | c | c | c | c |}
      \hline
      Data modality & Audio only & \multicolumn{2}{|c|}{EEG only} & Audio \& EEG \\ \hline
      Model    & PCA-SVM & DenseNet & PCA-SVM & PCA-SVM \\ \hline
      Accuracy & 67\%    & 61\%     & 59\%    & 81\%    \\ \hline
    \end{tabular}
  \end{center}
\end{table}
\subsection{Results on correlation learning between audio and EEG}

We use the cross-modal retrieval tasks to evaluate correlation between audio and EEG. We try to find the relevant audio for a given EEG signal and vice versa. In the former task, there is only one relevant audio, and the performance is measured by the MRR1 metric (the mean of the inverse of the rank of the relevant item), while the performance of the latter is measured by the MAP metric (there are 45 relevant EEG data for each audio, based on which mean average precision is computed). In the correlation analysis, we compare CCA \cite{Hotelling36}, DCCA \cite{And13} and C-DCCA \cite{yiyu18}, and their results are summarized in Tables \ref{tab:res_retrieval_2} and \ref{tab:res_retrieval}. According to these results, the retrieval performance is almost irrelevant of the number of CCA components, indicating that 10 CCA components are sufficient to achieve a good performance. CCA and DCCA produce similar results. In the task of retrieving audio from EEG, 720 EEGs (queries) correspond to 16 audios (in the database), and many EEG signals (45) share the same audio. This is equivalent to the classification of EEG to one of the 16 audios, and MRR1 is relatively high. In contrast, given an audio as query, there are 45 relevant EEG data in the database. But not all of them are quite similar even in the shared canonical space. Therefore, the MAP performance is relatively low in all methods. In both tasks, C-DCCA achieves a much better performance than the other two methods by stressing the intra-class similarity.



\begin {table}[h]
\caption {MRR1 of audio retrieval with EEG data as query} \label{tab:res_retrieval_2}
  \begin{center}
    \begin{tabular}{ | c | c | c | c |}
      \hline
      Number of components & CCA & DCCA & C-DCCA \\ \hline
      10 & 0.287 & 0.284 & 0.369 \\ \hline
      15 & 0.289 & 0.248 & 0.372 \\ \hline
      20 & 0.283 & 0.267 & 0.373 \\ \hline
      25 & 0.283 & 0.251 & 0.370 \\ \hline
      30 & 0.283 & 0.288 & 0.368 \\ \hline
      35 & 0.282 & 0.283 & 0.368 \\ \hline
      40 & 0.283 & 0.273 & 0.370 \\ \hline

    \end{tabular}
  \end{center}
  \end{table}

\begin {table}[h]
\caption {MAP of EEG retrieval with audio as query} \label{tab:res_retrieval}
  \begin{center}
    \begin{tabular}{ | c | c | c | c |}
      \hline
      Number of components & CCA & DCCA & C-DCCA \\ \hline
      10 & 0.112 & 0.115 & 0.182 \\ \hline
      15 & 0.112 & 0.092 & 0.182 \\ \hline
      20 & 0.109 & 0.105 & 0.188 \\ \hline
      25 & 0.108 & 0.099 & 0.186 \\ \hline
      30 & 0.111 & 0.129 & 0.184 \\ \hline
      35 & 0.109 & 0.109 & 0.183 \\ \hline
      40 & 0.109 & 0.114 & 0.183 \\ \hline

    \end{tabular}
  \end{center}
\end{table}

\section{Conclusion and Future work}

Experimental results confirm that using EEG helps to increase the precision of audio event classification. Meanwhile, the great gap between testing and training in the accuracy shows that we could increase the performances of all the classifiers by adding some regularization to avoid overfitting. We also notice that the detected correlation remains weak and further experiments or data collection are necessary to show more meaningful results. In the future, we will also investigate how to leverage the power of human perception to refine audio event recommendation.

\section{Acknowledgement}

The authors would like to appreciate Francisco Raposo's help in EEG data collection and processing during his internship in NII. Many thanks go to volunteers who helped us to record EEG signals where they were listening to audio events.

\bibliographystyle{IEEEtran}
\bibliography{IEEEabrv,mybibfile}

\end{document}